# Medikal Alanda Kullanılan Giyilebilir Teknolojiler: Uygulamalar, Karşılaşılan Sorunlar ve Çözüm Önerileri
# Medical Wearable Technologies: Applications, Problems and Solutions

*Erkan Bostancı [1]*

[1] Bilgisayar Mühendisliği Bölümü
Ankara Üniversitesi
ebostanci@ankara.edu.tr

### Özetçe

*Bu makalede tıp alanında da giderek yaygın bir şekilde kullanılmaya başlayan giyilebilir teknolojiler üzerinde durulmuştur. Akıllı saatlerden, akıllı gözlüklere; elektronik tekstilden, veri eldivenlerine kadar birçok teknolojik ürün çeşitli hastalıkların teşhis ve tedavisinde önemli rol oynamaktadır. Bu teknolojiler ile karşımıza çıkacak yeni tehditler de üzerinde durulması gereken önemli bir konudur. Veri gizliliğinden, büyük veri sorununa kadar farklı tehditler hayatımıza giren bu yeni teknolojinin bir yan etkisi olarak karşımıza çıkmaktadır. Makalede giyilebilir teknolojilerin uygulama alanları, karşılaşılan sorunlar hem teknik hem de etik bir bakış açısıyla irdelenmiş ve olası tehditlere karşı çözüm önerileri sunulmuştur.*

**Anahtar Kelimeler** — *giyilebilir teknolojiler; medikal uygulamalar; tehditler ve çözümler.*

### Abstract

*The focus of this paper is on wearable technologies which are increasingly being employed in the medical field. From smart watches to smart glasses, from electronic textile to data gloves; several gadgets are playing important roles in diagnosis and treatment of various medical conditions. The threats posed by these technologies are another matter of concern that must be seriously taken into account. Numerous threats ranging from data privacy to big data problems are facing us as adverse effects of these technologies. The paper analyses the application areas and challenges of wearable technologies from a technical and ethical point of view and presents solutions to possible threats.*

**Keywords** — *wearable technologies; medical applications; threats and solutions.*

## 1. Giriş

Gelişen teknolojilerle birlikte, kullanıcıların zihinlerinde yerleşik bilgisayar algısı artık kendisini masaüstü bilgisayarlardan akıllı telefonlara, tabletlere ve en nihayetinde de giyilebilir teknolojilere bırakmaktadır. Kullanıcı farkındalığının da artmasına bağlı olarak geniş bir ürün yelpazesiyle pazarda önemli bir yer edinen bu teknolojinin giderek artan bir şekilde hayatlarımızda yer edinmesinin artık işten bile olmadığı yapılan pazar araştırmalarıyla ortaya konmaktadır. Örnek olarak, Juniper araştırma şirketinin raporuna [1] göre 2013 yılının sonunda 8 milyar dolar olarak verilen satış hasılatının 2019 yılına 53,2 milyar dolara ulaşması bekleniyor.

Her ne kadar genel olarak eğlence sektörü yukarıda bahsedilen büyük bütçeler içerisinde aslan payına sahip olsa da medikal alanda kullanılan giyilebilir teknolojilerin giderek daha büyük önem kazanması çok uzun bir bekleme gerektirmeyecek. Özellikle de Avrupa ülkelerinde yaşlanan nüfusun artmasına karşın bakım sağlayacak personel sayısının az olması, doktorların hastaların durumlarını hastane dışında da uzun süreler boyunca takip edebilme ihtiyacı [2] ve bunun sonucunda da evde yapılabilecek öz bakım faaliyetlerinin artması gibi önemli nedenlerin [3] yanı sıra bilinç düzeyi artan kullanıcıların kendi sağlık durumlarını takip etme istekleri bu teknolojilere olan ihtiyacı artırmaktadır.

Medikal anlamda genel olarak sağlık durumunu izleme, tanı koyma ve tedavi gibi ana başlıklar altına alabileceğimiz giyilebilir teknolojiler hem klinik (nabız, şeker seviyesi) hem de davranışsal (yürüme, merdiven çıkma) veri toplama amacıyla kullanılmakta [4] ve birçok farklı uygulama örneğiyle karşımıza çıkmaktadır. İlaçların alınacağı zamanları hatırlatıcı sistemlerden, diyabet hastalarına belirli zamanlarda insülin enjekte edebilen sistemlere [5]; biyokimyasal algılayıcılardan [3] motor hareketlerin takibi [6] ve güçlendirilmesinde [7] kullanılan sistemlere ve hatta Alzheimer hastaları için geliştirilen hatırlatıcı gözlüklere [8] kadar birçok alanda karşımıza çıkan uygulamalar [9], aslında yakın gelecekte daha fazla neler yapılabileceğine dair önemli ipuçları ortaya koymaktadır.

Gelişen teknolojilerin önemli uygulamalara olanak tanıması oldukça önemli olduğu gibi bir yandan da bu teknolojilerle birlikte karşımıza çıkan yeni sorunların bu yeni uygulamalarda kullanılan teknolojileri de etkileyebileceği üzerinde dikkatlice düşünülmesi gereken bir konudur 10. İnternete bağlı her sistemde karşılaşılabilecek saldırıların bu durumda ölümcül sonuçlara yol açabileceği göz ardı edilemeyeceği gibi birçok farklı algılayıcı tarafından oluşturulacak devasa bilgi birikiminin kişilerin mahremiyetiyle ters düşmeyecek şekilde ve etik değerler ışığında araştırma amaçlı nasıl kullanılacağı gibi sorular henüz tam olarak cevaplanmış değil.

Bu makalede medikal alanda kullanımları giderek yaygınlaşan giyilebilir teknolojilere örnekler verilerek günümüz teknolojisinin geldiği nokta ortaya konacaktır. Farklı giyilebilir teknolojilerin kullanıldığı uygulama alanları Bölüm 2'de anlatılırken, bu uygulamalarla birlikte ortaya çıkan yeni tehditler dolayısıyla karşılaşılan sorunlara Bölüm 3''te değinilecektir. Bölüm 4'te de bu sorunlara getirilebilecek çözüm önerileri sunulacak ve makale sonuçlandırılacaktır.

## 2. Uygulama Örnekleri

Kablosuz teknolojiler sayesinde algılayıcı boyutlarının da küçülmesinin (Mikro Elektro Mekanik Sistemler, MEMS) yanı sıra düşük güç gerektiren bağlantılar ve gömülü işletim sistemlerinin geliştirilmesi ile birlikte algılayıcı teknolojisi yeni bir boyut kazanmış ve artık vücudun farklı bölgelerinden gelen bilgileri toplayan algılayıcılardan oluşan ağlar oluşturulmaya başlanmıştır [9].

Yapılan çalışmaların büyük çoğunluğunun kalp hareketlerinin takibi üzerine olduğu göze çarpmaktadır [11]. Geliştirilen giyilebilir sistemler ile kalp ritmi ve oksijen seviyelerinin ölçülebilmesi sağlanmıştır. Wang v.d. [12] tarafından geliştirilen kulak memesine takılan ve çoklu LED lambalar ve fotodiyotlar kullanan PPG (Photoplethysmogram) sistemi ile parmaktan yapılan ölçümlere göre %10 daha güçlü sinyaller alınabildiği ve böylece çok hassas ölçümler yapılabildiği belirtilmiştir. Suzuki v.d. [13] geliştirilen SILMEE sisteminde ise yama şeklinde göğse yerleştirilebilen bir algılayıcıdan gelen EKG, nabız ve vücut sıcaklığı bilgileri batarya süresi olarak belirtilen 24 saat boyunca IEEE 802.15.6 Bluetooth bağlantısı üzerinden bir akıllı telefona aktarılabilmektedir. Yakın zamanda geliştirilen akıllı saatlerle de nabız ve oksimetri ölçümleri yapılabilmektedir [14].

Massachusetts Institute of Technology (MIT)'de geliştirilen MIThril sistemi [3] ile bir RS-232 ara yüzü ile kalp ritmi, kan şekeri, oksimetri ve karbondioksit seviyesi takibinin yanı sıra elektro-ensefalografi (EEG) ölçümleri yapılabilmektedir. Farklı algılayıcılardan gelen ham ölçümlerden makine öğrenme yöntemleri ile istatistiksel modeller oluşturulmuş ve bu veriler kullanılarak çıkarımlar yapılmasına olanak sağlanmıştır.

MIThril sisteminin Parkinson hastaları için hipokinezi ve diskinezi durumlarının takibi ve analizi için de kullanıldığı belirtilmektedir. Yine Parkinson hastalarındaki semptomları incelenmesi amacıyla geliştirilen farklı bir giyilebilir sistemdeki [6] eylemsizlik algılayıcılarından gelen veriler destek vektör makineleri aracılığıyla sınıflandırılarak diskinezi ve bradikinezi durumlarının değerlendirilmesi sağlamıştır.

Romatoid Artrit hastaları için geliştirilen ve bir veri eldiveni kullanan sistem [15] sayesinde hem teşhis yapılabilmekte hem de tedavide kullanılan egzersizlerin yapılma doğruluğu kontrol edilebilmektedir. Eldivenlerde bulunan fiber optik eğme algılayıcıları parmak hareketleri rahat bir şekilde modellenebilmektedir. Teşhis aşamasında eldeki şekil bozukluğunun ve parmak eklemlerindeki sertliği derecesinin yanında eklemlerdeki şişme ve parmaklardaki konumsal kayma miktarının ölçümü yapılabilirken; tedavi amaçlı kullanımda ise geliştirilen kullanıcı dostu ara yüz ile önerilen egzersizlerin hangi doğrulukta yapıldığı kontrol edilebilmektedir.

Elektronik tekstil üzerine yapılan çalışmalar da giyilebilir teknolojiyi büyük oranda desteklemektedir [9]. Park v.d. [16] tarafından geliştirilen akıllı tişörtte farklı algılayıcılar ve bunlardan gelen verileri kaydeden bir cihaz bulunmaktadır. Yakın zamanda Patel v.d. [17] tarafından hazırlanan derleme bu konuda birçok farklı uygulama örneği sunmaktadır. Benzer şekilde Kaliforniya Üniversitesi tarafından geliştirilen Electrozyme (yeni adı *biolinq*) [18] sisteminde elektrokimyasal bir dövme ile kullanıcının ter analizi yapılabilmekte ve elektrolit dengesi, laktat birikimi gibi ölçümler yapılabilmektedir.

Giyilebilir teknolojilerin tıbbi eğitim amaçlı kullanımı da öne çıkan konular arasında yer almaktadır. Goodwin v.d. [19] tarafından yapılan çalışmada Google Glass takan tıp öğrencilerine simüle edilmiş omuz takılması (distosi) ve ters doğum gibi acil obstetrik durumlar gösterilmiştir. Deneyler sonrasında yapılan anketlerde katılan öğrencilerin tümü benzer simülasyonların acil obstetrik bakım sürecini iyileştireceğini belirtmiştir. Başka bir çalışmada [20] ise tıp eğitimi sırasında yeni stajyerlerin karşılaşabildikleri acil durumlarda yardımcı olması için kullanılan bir Google Glass ile yanlış teşhislerin önlenmesi ve hasta güvenliğinin sağlanması amacıyla farklı senaryolar denenmiştir. Google ve Novartis Alcon ortaklığı tarafından geliştirilen akıllı kontak lens sayesinde [21] hem vücut şeker seviyesi invazif olmayan bir yöntemle ölçülebilecek hem de lensin otomatik odaklama özelliği sayesinde presbitlik sorununun önüne geçilecektir.

Ülkemizde de araştırmacıların dikkatini çekmeye başlayan bu konuda son derece önemli çalışmalar yürütülmektedir. Duru v.d. [22] tarafından yapılan çalışmada Zigbee kablosuz haberleşme teknolojisi kullanılarak kalp atım ritmi ve vücut sıcaklığı uzak bir bilgisayara gönderilerek bu alınan verinin görselleştirilmesi sağlanmıştır. Yapılan başka bir çalışmada [7] ise engellilerin bir ev ortamı içinde Microsoft Kinect algılayıcısının insan vücudunu takip edebilme özelliğini kullanarak belirli komutlar vermesi ve bu sayede lambaların yanması, sıcaklığın ayarlanması ve kapıların kapatılması gibi işlemleri rahat bir şekilde yapabilmeleri sağlanmıştır. Algılayıcı amaçlı kullanılmak üzere elektronik tekstiller geliştirilmesi üzerine yürütülen projeler [23] konunun önemi üzerindeki farkındalığı ortaya koymaktadır.

## 3. Karşılaşılan Sorunlar

Verilen örneklerde de görüldüğü üzere büyük bir ilgi uyandıran giyilebilir teknolojilerin kullanıldığı sağlık uygulamalarında bilgi güvenliğinin önemi son derece büyüktür ve böyle bir uygulama alanında karşılaşılabilecek güvenlik tehditlerinin de iyi bir şekilde irdelenip gerekli önlemlerin alınması gerekmektedir. Üstelik bu teknolojiler genel olarak da mobil sistemlerin (farklı donanımlar üzerinde çalışan aynı işletim sistemleri) bir türevi olarak karşımıza çıktığından aynı sorunlar daha geniş bir etki alanında daha büyük bir etki yaratabilecek bir duruma gelmiştir.

Veri üreten ve bir bağlantı üzerinden bu veriyi aktaran her sistem gibi giyilebilir teknolojilerin de karşılaştığı veri gizliliği, kötü amaçlı yazılımlar ve bağlantı bağımlılığı gibi üzerinde durulması gereken tehditler bu teknolojilerin zayıf yönleri olarak karşımıza çıkmaktadır [10]. Ayrıca günümüzün popüler sorunlarından olan büyük veri sorunu ve üretilen veriyi verimli bir şekilde işleyecek entegre araçların olmayışı da önemli sorunlar arasındadır [4].

### 3.1. Veri Gizliliği

Birçok uygulamada öne çıkan veri gizliliği giyilebilir teknolojilerin tıbbi alanda kullanılmasıyla yeni bir boyut kazanmaktadır. Klinik muayene veya teşhis esnasında özen gösterilen hastanın mahremiyet hakkına saygı ve tıbbi değerlendirmenin gizlilik içerisinde yürütülmesi [24] gibi hassasiyet gerektiren ilkelerin özellikle de aşağıdaki bölümlerde bahsedilecek tehditlerden dolayı bu yeni teknolojilerin kullanımı sırasında da göz ardı edilmemesi gerekmektedir.

Bahsedilecek Internet tabanlı saldırıların yanı sıra örnek olarak, son derece masumane niyetlerle, nabız ölçümü yapan bir giyilebilir teknoloji kullanması önerilen bir hastanın özel hayatının doktoru tarafından takip edilebileceğinin bilinci hasta üzerinde nasıl bir etki oluşturacağının düşünülmesi [25], bu durumun hastanın farkındalığı olmadığı durumlarda doktor tarafından takip edilmesinin doğruluğu irdelenmelidir.

### 3.2. Kötü Amaçlı Yazılımlar

Trojanlardan istenmeyen reklamlara kadar geniş bir yelpazede [26] incelenebilen kötü amaçlı yazılımlar mobil teknolojilerle birlikte benzer işletim sistemleri (Android, IOS) üzerine çalışan giyilebilir teknolojileri tehdit etmektedir. Hedef sistemden bilgi çalma, belirli hizmetlerden faydalanmasını engelleme (sunucuyu meşgul ederek) gibi faaliyetlerde bulunan kötü amaçlı yazılımların ne tür saldırılar için kullanılacağı aslında bilgisayar korsanlarının hayal gücüne kalmıştır.

Jay Freedman tablet bilgisayarlar ve cep telefonları için geliştirdiği yazılım ile Android tabanlı Google Glass üzerinde yönetici izinlerine sahip olmuştur [27] ve bunu arkadaşları ile akşam yemeği yerken iki saat içinde yapabildiğini belirtmiştir. Bu şekilde saldırılan bir akıllı gözlük ile yürütülen bir tıbbi operasyon düşünüldüğünde, akıllı gözlüğün kamerası ile çekilen hastalara ait görüntülerin izinsiz olarak İnternette yayılması işten bile değildir. Kötü amaçlı yazılımlarla giyilebilir teknolojiler üzerinden ne tür saldırılar yapılabileceğine dair daha birçok örnek verilebilir.

### 3.3. Bağlantı Bağımlılığı

Veri aktarımı, güncelleme gibi önemli nedenlerle kablosuz olarak İnternete bağlanan giyilebilir teknolojiler verimlilikleri artırmak için kullandıkları bu teknoloji üzerinden yeni saldırılara da kendilerini açık bir hale getirmektedir. Güvensiz bağlantılar kullanıldığında maruz kalınan iki tür saldırı türü bulunmaktadır: aradaki adam (man-in-the-middle) ve paket koklama (sniffing).

Aradaki adam türünde saldırılarda saldırgan birbirleri ile direk olan haberleştiklerini düşünen iki iletişim uç noktası arasına girer ve büyük çoğunlukla da bu iletişimde aktarılan veriyi değiştirir. Koklama türü saldırılarda da ağ üzerinde gönderilen paketler saldırgan tarafından incelenir ve bu sayede şifreleme ile korunmayan hassas bilgiler ele geçirebilir.

### 3.4. Büyük Veri Sorunu

Büyük veri sorununu, günümüzde kullanılan teknolojiler tarafından üretilen veri ile artık klasik veri işleme süreçleri (veri toplama, arama, gizlilik, görselleştirme, analiz vs.) ile baş edilememesi olarak tanımlayabiliriz. Teknolojinin kullanıldığı hemen her alanda karşımıza çıkmaya başlayan büyük veri sorunu tıp alanındaki uygulamalarda da kendi göstermektedir.

Çok sayıdaki giyilebilir teknolojiler tarafından her an üretilen büyük miktarlardaki verinin güvenliği ve gizliliği de göz önünde bulundurularak toplanıp, saklanmasına ve incelenebilmesine olanak tanıyacak sistemler gerekmektedir. Toplanan verilerden elde edilen çıkarımların doktorlar tarafından önerilecek tedavi yöntemlerinde yol gösterici olması beklenmektedir.

### 3.5. Analiz Araçlarının Uyumsuzluğu

Giyilebilir teknolojiler tarafından üretilen verilerin doktorlar tarafından incelenerek yorumlanması ve bu verilerin katma değerinin araştırılması gerekmektedir. Kullanılan veri formatlarında bir standartlaşmanın yetersiz ölçüde oluşu veya olmayışı bu verilerin incelenmesinde kullanılacak entegre sistemlerin geliştirilmesinde bir sorun oluşturmaktadır.

Bir diğer sorun ise geliştirilecek entegre sistemlerin hali hazırda bulunan otomasyon sistemleriyle uygun bir şekilde çalışması gerektiğidir. Hasta özlük bilgileri ile giyilebilir teknolojilerden gelen verinin entegre bir sistem içerisinde doktor tarafından görülebilmesi ve muayene sırasında konulan teşhisin de bu sistemde iyi bir şekilde harmanlanabilmesi gerekmektedir.

## 4. Çözüm Önerileri ve Sonuçlar

Nabız ve vücut ateşi ölçen sistemlerden, hatırlatıcı gözlüklere; rehabilitasyon amaçlı kullanılan akıllı tekstillerden akıllı dövmelere kadar geniş bir yelpazede ürünlerle karşımıza çıkan giyilebilir teknolojilerin yakın gelecekte hayatımızda daha geniş bir yer tutması artık uzak değil. Bu teknolojilerin tıp alanında da kullanılmaya başlanmasıyla klinik dışı teşhis kolaylaşacak, hasta öz bakımı desteklenecek, tıbbi eğitim faaliyetleri desteklenecek ve böylece giderler azalacak ve olası yanlış teşhis sonucu oluşabilecek hasta mağduriyetinin de önüne geçilecektir.

Birçok faydasının yanında bu teknolojilerin daha yaygın kullanımıyla karşılaşılabilecek veri gizliliği, kötü amaçlı yazılımlar, oluşan büyük miktarlardaki verinin işlenmesi ve var olan sistemlerle entegrasyon gibi tehditlerin de üzerinde durulması ve bu tehditlerin giyilebilir teknolojinin yararlarının önüne geçmesi önlenmesi gerekmektedir.

Karşılaşılan veya ortaya çıkması öngörülen sorunlar için de aşağıdaki önlemler alınarak tehditler en aza indirgenebilir:

- Giyilebilir teknolojilerde kullanılan işletim sistemlerinde diğer mobil cihazlarda kullanılanlarla aynı değil fakat bunların özelleştirilmiş versiyonlarının kullanılması fayda sağlayacaktır. Bu sayede cihazlarda yeni yazılım yükleme, güvensiz bağlantılara izin verme veya tıbbi veri toplayan programın dışında herhangi bir program çalıştırmama gibi yetki kısıtlamaları yapılabilir.
- Birçok farklı sistemin üreteceği tıbbi veriler için yeni bir standart biçim geliştirilmesi gerekebilir. Bu sayede kullanılan verinin homojenliği artırılarak farklı sistemlerin entegrasyonu ortak veri biçimi üzerinde yapılabilir.
- Oluşacak büyük miktarlardaki verilerin işlenmesi (büyük veri sorunu) için de özelleştirilmiş sunucular üzerinde yapılacak paralel hesaplamalar yol gösterici olacaktır. Özellikle de grafik kartları üzerinden hesaplama yapmaya olanak tanıyan teknolojiler ile bu darboğaz aşılabilir.
- Veri gizliliği ile ilgili olarak da gizliliği sağlayacak teknik altyapının sağlanmasının (şifreleme protokolleri) yanı sıra

ilgili mevzuatın, bu teknolojilerin getirebileceği yeni durumlar da göz önünde bulundurularak, gözden geçirilmesi ve yeni maddeler ile güncellenmesi gerekebilir.

Tıbbi uygulamalarda büyük faydalar sağlayacağı öngörülen giyilebilir teknolojiler, bu teknolojilerin kullanımında ortaya çıkabilecek tehdit unsurlarının da önüne geçilmesiyle, daha sağlıklı bireylerden oluşacak bir sağlıklı ve mutlu bir toplum olmamız yolunda bizlere önemli katkılar sağlayacaktır.

## 5. Kaynakça